\documentclass[12pt]{article}
\usepackage{array}
\usepackage{graphicx,epsfig}
\usepackage{amsmath,amssymb,amsthm,bm}
\usepackage{verbatim}
\usepackage{subfigure}
\usepackage[english]{babel}
\usepackage{latexsym,psfrag,array,multicol,palatino,enumerate,caption,multirow}
\usepackage{setspace}
\usepackage{fullpage}
\usepackage[usenames]{color}
\usepackage{bbm}
\usepackage{arydshln}
\usepackage[square,numbers]{natbib}
\setcitestyle{numbers}
\usepackage{authblk}

\usepackage{hyperref}
\hypersetup{
     colorlinks   = true,
     linkcolor    = blue,
     citecolor    = blue
}

\definecolor{DarkBlue}{rgb}{0.1,0.1,0.5}
\definecolor{Red}{rgb}{0.9,0.0,0.1}
\definecolor{violet}{rgb}{0.8,0.0,0.2}

\usepackage{dsfont}

\usepackage{multirow}

\newcommand\blfootnote[1]{%
  \begingroup
  \renewcommand\thefootnote{}\footnote{#1}%
  \addtocounter{footnote}{-1}%
  \endgroup
}

\usepackage[margin=1in]{geometry}

\makeatletter
\def\ps@pprintTitle{%
  \let\@oddhead\@empty
  \let\@evenhead\@empty
  \let\@oddfoot\@empty
  \let\@evenfoot\@oddfoot
}
\makeatother

\begin{document}

\title{High-Dimensional Directional Brain Network Analysis for Focal Epileptic Seizures}

\author[a,1]{Yaotian Wang}
\author[b]{Guofen Yan}
\author[c]{Seiji Tanabe}
\author[d]{Chang-Chia Liu}
\author[d]{Shayan Moosa}
\author[e,1]{Mark S. Quigg}
\author[a]{Tingting Zhang}

\affil[a]{Department of Statistics, University of Pittsburgh, Pittsburgh, PA 15206}
\affil[b]{Department of Public Health Sciences, University of Virginia, Charlottesville, VA 22904}
\affil[c]{Department of Psychology, University of Virginia, Charlottesville, VA 22904}
\affil[d]{Department of Neurosurgery, University of Virginia, Charlottesville, VA 22904}
\affil[e]{Department of Neurology, University of Virginia, Charlottesville, VA 22904}

\date{}

\clearpage\maketitle
\thispagestyle{empty}

\blfootnote{$^1$Corresponding authors. To whom method questions should be addressed: Yaotian Wang, E-mail address: YAW87@pitt.edu. To whom data questions should be addressed: Mark S. Quigg, E-mail: MSQ6G@hscmail.mcc.virginia.edu.}

\begin{abstract}
The brain is a high-dimensional directional network system consisting of many regions as network nodes that influence each other. The directional influence from one region to another is referred to as directional connectivity. Epilepsy is a directional network disorder, as epileptic activity spreads from a seizure onset zone (SOZ) to many other regions after seizure onset. However, directional network studies of epilepsy have been mostly limited to low-dimensional directional networks between the SOZ and contiguous regions due to the lack of efficient methods for analyzing high-dimensional directional brain networks. To address this gap, we study high-dimensional directional networks in epileptic brains by using a clustering-enabled multivariate autoregressive state-space model (MARSS) to analyze multi-channel intracranial EEG recordings of focal seizures. This new MARSS characterizes the SOZ, nearby regions, and many other non-SOZ regions as one integrated high-dimensional directional network system with a clustering feature. Using the new MARSS, we reveal changes in high-dimensional directional brain networks throughout seizure development. We simultaneously identify directional connections and the SOZ cluster, regions most affected by SOZ activity, in different seizure periods. We found that, after seizure onset, the numbers of directional connections of the SOZ and regions in the SOZ cluster increase significantly. We also reveal that many regions outside the SOZ cluster have no changes in directional connections, although these regions' EEG data signal ictal activity. Lastly, we use these high-dimensional network results to localize the SOZ and achieve 100\% true positive rates and $\leq$3\% false positive rates for different SOZ locations.
\end{abstract}\hspace{10pt}


\section{Introduction}
The brain is a high-dimensional directional network system consisting of many regions as network nodes that exert influences onto each other. The directional influence from one brain region to another is referred to as directional connectivity (or effective connectivity \citep{FristonBook04}) and corresponds to one directional edge in a directional brain network. The high-dimensional directional network refers to not only the many nodes in the network but also the dramatic increase in potential directional edges between the nodes as the number of the nodes increases (typically, $>$30 nodes).  In epilepsy, the propagation of seizures from a seizure onset zone (SOZ) to other healthy brain regions is a clinically important example of pathological directional brain networks \citep{bernhardt2015network,englot2016regional}. However, existing directional brain network results of epilepsy were mostly about low-dimensional directional networks between the SOZ and a few contiguous regions  \citep{korzeniewska2014ictal} or on functional connectivity without directionality information \citep{lee2014altered,stacey2019emerging}. Results of high-dimensional directional networks between the SOZ, adjacent regions, and many other non-SOZ regions are lacking. Extending SOZ network studies to include many other non-SOZ regions can better elucidate the pathological mechanism and the consequences of seizure propagation. Therefore, our study aimed to characterize the SOZ, adjacent regions, and many other non-SOZ regions as one integrated high-dimensional directional network system and reveal changes in this directional network during the transition from interictal to ictal phases.

Existing network methods have limitations in identifying high-dimensional directional networks in the brain. For example, pairwise information theoretic measures for directional connections, such as transfer entropy \citep{Sabesan09,Schreiber00,vicente2011transfer} and directed transinformation \cite{Hinrichs06,Saito81}, fail to separate direct and indirect connections (i.e., the connections mediated by other regions). Most model-based approaches were developed mainly for low-dimensional directional networks. These approaches include dynamic causal modeling (DCM)
\citep{David06,Friston03,Kiebel06,Kiebel08,Marreiros10,Penny04}, structure equation modeling \citep{astolfi2004estimation,McIntosh94}, Granger causality \citep{marinazzo2011nonlinear}, multivariate autoregression \citep{Goebel03,Roebroeck05}, multivariate autoregression-based partial directed coherence \citep{baccala2001partial}, Gaussian Bayesian network \citep{wu2011altered,zheng2006learning}, and many others \citep{sakkalis2011review}. Using these approaches to analyze high-dimensional brain networks is susceptible to large estimation errors. Despite efforts of extending these models to high-dimensional directional network analysis, such as spectral DCM \citep{friston2014dcm,razi2017large}, sparse regression DCM \citep{frassle2018generative}, and other high-dimensional ordinary differential equation models \citep{Zhang14DDM,Zhang15BDDM,Zhang17BMODDM}, these models still have large estimation errors and are computationally intensive.

We recognize three main challenges that require overcoming in the development of an effective modeling approach to identify high-dimensional directional networks in the brain: i) high computational intensity, ii) model inadequacy, i.e., a deviation of the assumed model from the underlying true brain network model, and iii) low estimation efficiency of many parameters in the model. We use a new modeling approach \citep{li2020bayesian} to address these challenges. This new approach integrates a multivariate autoregressive state space model (MARSS) with a clustering feature (also called modularity \citep{Newman06}) to characterize both directional connections and clusters in high-dimensional directional brain networks.

 The MARSS consists of two sub-models, a state model and an observational model. The state model characterizes directional connections between regional brain activities, while the observational model links observed brain data to the regional brain activities. Both sub-models are in the form of linear equations, which enables computational feasibility. Each sub-model has its own error term: the model error in the state model and the data measurement error in the observation model. These two separate errors make the MARSS robust to model inadequacy and data noise. To enhance estimation efficiency for many parameters in the MARSS and produce scientifically meaningful network results, we incorporate the clustering feature in the MARSS as a prior for the underlying brain network. We develop a Bayesian framework to estimate this new MARSS and simultaneously identify directional connections and clusters. Overall, under the Bayesian framework, the new MARSS with the clustering feature offers
computational feasibility, improved estimation efficiency, and robustness to model inadequacy and data
noise for identifying high-dimensional directional networks in the brain.

We apply the new MARSS to multi-channel intracranial EEG recordings from 6 patients with focal epilepsy. The following first introduces the new MARSS used in this study and then presents our findings regarding the high-dimensional directional networks in epileptic brains. We show the evolution of these patients' high-dimensional directional networks from interictal to ictal phases. Changes in directional connections and clusters are uncovered not only for the SOZ and adjacent regions but also for many more distant non-SOZ regions. We also use these network results to localize the SOZ independently from traditional visual analysis in the clinical practice to demonstrate the effectiveness of our method in uncovering different directional connectivity properties of different regions. We compare identified candidate SOZ regions against patients' clinically localized SOZ (used as the given truth).  We show that our method achieved high accuracies in localizing the SOZ (100\% true positive rates and less than 3\% false positive rates) for all 6 patients.

\section{Materials and Methods}\label{sec:method}

\subsection{The High-Dimensional Directional Network Model}
Let $\mathbf{y}(t)=(y_1(t),\ldots,y_d(t))^\prime$ be a $d$-channel intracranial EEG recordings of $d$ regions at time $t$ and $\mathbf{x}(t)=(x_1(t),\ldots,x_d(t))^\prime$ be the brain activities of the $d$ regions at $t$. The MARSS consists of two sub-models with the following forms
\begin{align}
&\mbox{Observation model: }y_i(t)= c_i\cdot x_i(t) +  \epsilon_i(t), \label{eq:obseq}\\
&\mbox{State model: }x_{i}(t) = \sum_{j=1}^{d} \gamma_{ij} \cdot A_{ij} \cdot x_{j}(t-1) + \eta_i(t), i = 1, \ldots, d, ~\mbox{and}~ t = 1, \ldots, T,\label{eq:SBSSVAR}
\end{align}
where $c_i$ is a parameter for standardizing activities of different regions; $\gamma_{ij}$ is an indicator with 1 indicating the presence of the directional connection from region $j$ to region $i$ in the directional network and with 0 for the absence; $A_{ij}$s are coefficients; and $\eta_i(t)$ and $ \epsilon_i(t)$ are error terms with mean zero.

The observation model, equation \eqref{eq:obseq}, links the observed intracranial EEG recordings, $\mathbf{y}(t)$, to underlying brain activities of $d$ regions, $\mathbf{x}(t)$. The error $ \epsilon_i(t)$ is the data measurement error for brain region $i$.  The state model, equation \eqref{eq:SBSSVAR}, describes directional connections between $d$ regions' activities: Region $j$ has a directional influence over $i$, if and only if $\gamma_{ij}=1$. The error $\eta_i(t)$ is the model error, representing the deviation of the assumed linear equation from the underlying true model.  We use indicators $\gamma_{ij}$s in the MARSS to distinguish nonzero directional connections from zero ones. Under the MARSS, our focus is on selecting nonzero $\gamma_{ij}$s.

Note that the MARSS is not used to explain variation of brain activity at different regions, in contrast to other models, such as the DCM \citep{David06,Friston03,Kiebel06,Kiebel08,Marreiros10,Penny04}. Instead, we use the MARSS as a working model to identify directional connections through detecting existence of temporal dependence among regions' temporal activity. Though it is possible that more complex models, such as high-order multivariate autoregression, can fit the data better, however, these models also contain significantly more parameters, and the ensuing parameter estimates have much larger estimation errors. Moreover, the goal here is not to find the exact order of temporal dependence among regional brain activities or produce a perfect model that can fit all regions' activities. Overall, we use the first-order MARSS, with model sparsity, flexibility, and robustness, to capture the existence of temporal dependence among regions' activities.

\subsection{Bayesian Hierarchical Model for the Clustering Structure}
Despite the advantages of the MARSS, equations \eqref{eq:obseq} and \eqref{eq:SBSSVAR}, in robustness and computational feasibility, when used for high-dimensional directional networks, it still has low efficiency in parameter estimation (i.e., large estimation errors) due to many parameters in the MARSS when $d$ is large. We improve estimation efficiency by incorporating the clustering feature into the MARSS estimation. The clustering feature is a network pattern in which network nodes within the same cluster interact more strongly and densely with each other than with nodes in different clusters. The clustering feature has been widely reported in the literature on brain networks \citep{Milo02, Milo04, Newman06,Sporns11}. Thus, brain network estimates with the clustering feature are consistent with existing knowledge of the brain's functional organization.

Incorporating the clustering feature in a statistical model for a high-dimensional directional brain network and using the model to identify clusters and directional connections in the network are not straightforward. Note that the clustering feature is a special form of sparse networks, because it indicates that connections between clusters are sparse. However, a general sparsity regularization \citep{basu2015regularized} for brain network estimates \citep{frassle2018generative,Valdes05}, which treats every region equally, does not have the clustering feature and cannot detect clusters. Most existing approaches \citep{sporns2016modular,zalesky2012connectivity} to cluster identification are based on known connections between regions (i.e., known network edges). In summary, most existing approaches identify clusters and connections in separate steps using different methods, which can result in two different errors in the estimated network. In contrast, our new method simultaneously identifies clusters and directional connections for the brain networks whose network edges (presence or absence) are unobserved.

To enable clustering of regions in the model for an unobserved network, we propose a stochastic blockmodel prior for the unknown MARSS parameters that represent directional connections. The standard stochastic blockmodel \citep{airoldi2008mixed,durante2014nonparametric,fienberg1985statistical,holland1983stochastic,lorrain1971structural,nowicki2001estimation}
is a generative model for networks \citep{arroyo2017network,karrer11,paul2018random,rohe11} that have the clustering feature. We here modify the stochastic blockmodel to be a prior for the MARSS parameters $\gamma_{ij}$s, as explained below.

Let $K$ be the pre-specified number of clusters in the brain network. Let $\bm{m}_i=(m_{i1},\ldots, m_{iK})^\prime$ be a $K$-dimensional vector with only one element being 1 and the rest being 0, which labels the cluster of region $i$. For example, $m_{ik}=1$ indicates region $i$ is in cluster $k$. Let $\mathbf{B}=\{B_{k_1k_2},k_1,k_2=1,\ldots,K\}$ be a $K\times K$ matrix whose element $B_{k_1k_2}$ takes a value between 0 and 1 and denotes the prior probability of a nonzero directional connection from a region in cluster $k_2$ to another region in cluster $k_1$.

The prior distribution for the clustering feature is a joint distribution for indicators $\gamma_{ij}$s, the cluster labels $\bm{m}_i$s, and the probability matrix $\mathbf{B}$ as follows:\vspace{-0.1cm}
\begin{eqnarray}
&\gamma_{ij}|\bm{m}_i,\bm{m}_j,\mathbf{B}\stackrel{\mbox{ind}}{\sim} \mbox{Bernoulli}(\bm{m}_i^\prime~ \mathbf{B}~ \bm{m}_j);\label{eq:priorIndicator}\\
& \bm{m}_i\stackrel{\mbox{i.i.d}}{\sim} \mbox{Multinomial}(1;p_1,\ldots,p_K), ~i=1,\ldots, d,~~\mbox{and}~~
(p_1,\ldots,p_K)\sim \mbox{Dirichlet}(\bm{1_K});\label{eq:Multi-Dirichlet}\\
&B_{kk} \stackrel{\mbox{i.i.d}}{\sim} \mbox{Uniform}(l_0, 1), ~
B_{k_1k_2} \stackrel{\mbox{i.i.d}}{\sim} \mbox{Uniform}(0, u_0), ~k,k_1,k_2 = 1, \ldots , K,~ k_1 \neq k_2; \label{eq:priorProb}
\end{eqnarray}
where $l_0$ and $u_0$ are set at 0.9 and 0.1 to reflect dense connections within clusters and sparse connections between clusters \citep{ParkFriston2013}, and $\bm{1_K}$ is a vector consisting of $K$ 1s.

The distributions \eqref{eq:priorIndicator}, \eqref{eq:Multi-Dirichlet}, and \eqref{eq:priorProb} together define the stochastic blockmodel prior. We assign standard non-informative priors (e.g., almost flat Gaussian and inverse gamma distributions) to the remaining parameters.

We use the Gibbs sampler to simulate the proposed hierarchical Bayesian model, i.e., \eqref{eq:obseq}-\eqref{eq:priorProb}, and priors for other parameters. See details of the Markov chain Monte Carlo (MCMC) simulations in \cite{li2020bayesian}.

Let $\theta^{(s)}$ denote the $s$th posterior sample of the parameter $\theta$ after burn-in. Through posterior simulations, we obtain two posterior probabilities for every pair of regions $i$ and $j$: the posterior probability that the directional connection from region $j$ to region $i$ is nonzero, $\sum_{s=1}^S \gamma^{(s)}_{ij}/S$; and the posterior probability that regions $i$ and $j$ are in the same cluster, $\sum_{s=1}^S \mathbbm{1}({\bm{m}^{(s)}_i=\bm{m}^{(s)}_j})/S$, where $\mathbbm{1}(\cdot)$ is an indicator function, and $S$ is the total number of Gibbs sampler iterations after burn-in.

\subsection{Subjects and intracranial EEG Recordings}
We analyzed intracranial EEG recordings of 6 patients with medically intractable focal seizures who underwent evaluation for epilepsy surgery. Intracranial EEG implantation was customized for each patient. Each patient had at least 3 clinical and electrographic seizures with clear SOZs. A board-certified EEG expert examined intracranial EEG data and determined SOZs and seizure onset times. Table \ref{table:patients} presents information about 6 patients with drug-resistant epilepsy in this study. The information includes the number of seizures analyzed, the age of the patient at the recording, epilepsy types, etiology, and surgical outcome for each patient.

\begin{table}
\caption{Patient information. \label{table:patients}}
\centering
\begin{tabular}{c c  cc c c  c} 
\hline\hline
Patient  & Number of &  Age & Gender & Epilepsy          & Etiology  &Surgical\\
         & Seizures&      &        &                   &  &\\
         & Analyzed&      &        &  Type             &           &Outcome\\
\hline
1        &3    &  23    &M  &left lateral&dysplasia      & EC2 (No\\
         &     &        &   & frontal    &               & surgery, but  \\
         &     &        &    &                       &           &RNS installed)\\    \hdashline
2        &7    &  57    &M    &right       &gliosis   &EC1B\\
         &     &        &     &  temporal  &          &   \\    \hdashline
3        &4   &  28    &F    &left inferior&dysplasia&EC1B\\
         &     &        &     &  frontal    &         &     \\    \hdashline
4             &  4     &  30 &M    &left temporal&dysplasia&EC1A\\
              &        &     &     &parietal     &         &    \\    \hdashline

5        &  5        &  44    &M    &left lateral&dysplasia &EC1A\\
         &           &         &    & frontal     &          &    \\    \hdashline
6        &  3        &  45    &M    &right       &unk       &EC3\\
         &           &       &     &  subtemporal &        &\\  \hline
\end{tabular}\\
\footnotesize{ Age indicates the patient's age at recording.}
\end{table}

For each patient, we analyzed his/her intracranial EEG recordings from 300 seconds before to 300 seconds after electrographic seizure onset. The recordings were down-sampled to 1,024 Hz (collected at a 4,096 Hz [Natus, Middleton, WI USA]), filtered with a 60 Hz notch filter, and underwent removal of the first principal component to minimize artifacts. The analysis was blinded to the location of clinically determined SOZ but not the seizure onset time so as to provide a common reference time point with respect to the starting time of electrographic ictal activity. Identified directional brain networks (averaged across seizures) were presented for every 25-second window.
\subsection{Identification of directional brain networks}
For each patient, we identified his/her directional brain network in any 25-second window through simultaneously identifying directional connections and clusters in the network, using the above two probabilities generated from the analysis of intracranial EEG data in the window, as explained in detail below.

We first used the new MARSS to analyze the patient's 1-second intracranial EEG segments independently and obtained the above two posterior probabilities for every 1-second segment. For every pair of regions $i$ and $j$, we evaluated the evidence for the directional connection from region $j$ to $i$ being present in the window from $t$ to $t+25$ seconds (the $[t,t+25]$ window) by averaging the posterior probabilities of $\gamma_{ij}=1$ from all non-overlapping 1-second segments in the window across all seizures, where $t=0$ is the seizure onset time, and $t=-300,-299,\ldots,275$. We denoted the ensuing average probability by $P^{t,d}_{ij}$ and referred to it as the network edge probability. The network edge probability, $P^{t,d}_{ij}$, quantifies how likely, on average, the directional connection from region $j$ to $i$ was present in the 25-second window. Similarly, we obtained the average posterior probability of two regions, $i$ and $j$, in the same cluster in the window. This average probability is denoted by $P^{t,c}_{ij}$ and referred to as the clustering probability. The clustering probability, $P^{t,c}_{ij}$, quantifies how likely, on average, regions $j$ and $i$ were in the same cluster in the $[t,t+25]$ window.

For each 25-second window $[t,t+25]$, we identified its directional brain network by using network edge probabilities, $P^{t,d}_{ij}$s, to identify directional connections and by using clustering probabilities, $P^{t,c}_{ij}$s, to identify clusters.  For every pair of regions, $i$ and $j$, if $P^{t,d}_{ij}$ was greater than a threshold, the directional connection from region $j$ to $i$ was deemed to be present in that window. Similarly, if $P^{t,c}_{ij}$ was greater than a threshold, regions $j$ and $i$ were deemed to be in the same cluster in that window. To determine the thresholds, we used the earliest 25-second window (275-300 seconds before seizure onset time in our study) as the baseline window. The top 1\% percentile of all the $P^{-300,d}_{ij}$s in the baseline window was used as the threshold for network edge probabilities (in all windows), and the top 1\% percentile of all the $P^{-300,c}_{ij}$s in the baseline window was used as the threshold for clustering probabilities.  We used the top 1\% percentiles as the thresholds to select the most significant connections and to keep a low false discovery rate. In addition, top 1\% percentiles have been commonly used as thresholds in Bayesian variable selection problems \citep{Zhang14Bayes,mccann2007robust}.

\subsection{Quantification of extent of directional connectivity}
For each region $j$, we quantified its extent of directional connectivity in the window from $t$ to $t+25$ seconds by averaging region $j$'s all the network edge probabilities in the window: $$D_j^t=\sum_{i=1}^d (P^{t,d}_{ij}+P^{t,d}_{ji})/2d,$$ where $d$ is the total number of regions in the MARSS model.

We then examined the change in the extent of directional connectivity for region $j$ at time $t$ by $$DC^t_j=D^{t}_j-D^{t-25}_j.$$ That is, $DC^t_j$ is the difference between the average directional connectivity in the two windows that overlap at the time point $t$ only.
We evaluated $DC^t_j$ at every one second $t$. We refer to $DC^t_j$ as the directional connectivity change of region $j$ at time $t$.

\subsection{SOZ localization}
To assess the effectiveness of our method in revealing different regions' directional connectivity properties, we proposed to localize the SOZ based on $DC^t_j$. Let $t=0$ denote the seizure onset time. Let $M_j=\max\{DC^t_j, t\leq 0\}$, the maximum directional connectivity change of region $j$ over time no later than the seizure onset time. Let $C_{0}$ be the top 10\% percentile of $\{DC^0_j, j=1,\ldots,d\}$, the directional connectivity changes of all the regions at $t=0$. A region $l$ was selected to be a candidate for the SOZ if $DC^0_l$, its directional connectivity change at $t=0$, was greater than both $0.9M_l$ and $C_{0}$.

\section{Results}

\subsection{Simulations}

 This section presents simulation studies of the new MARSS, assessing its robustness and efficiency in identifying
directional connections for data generated from models distinct from our MARSS. We used the dynamic causal modeling (DCM) \citep{Friston03,kiebel2008dynamic} to simulate data because the DCM is the most popular model for directional connectivity.

\begin{figure}[h]
\centering
\subfigure[\footnotesize{Simulated Network DCM-EEG}]{
\label{fig:sys5}
    \includegraphics[width=5.2cm,height=4.00cm,trim= 10mm 30mm 10mm 30mm,clip=TRUE]{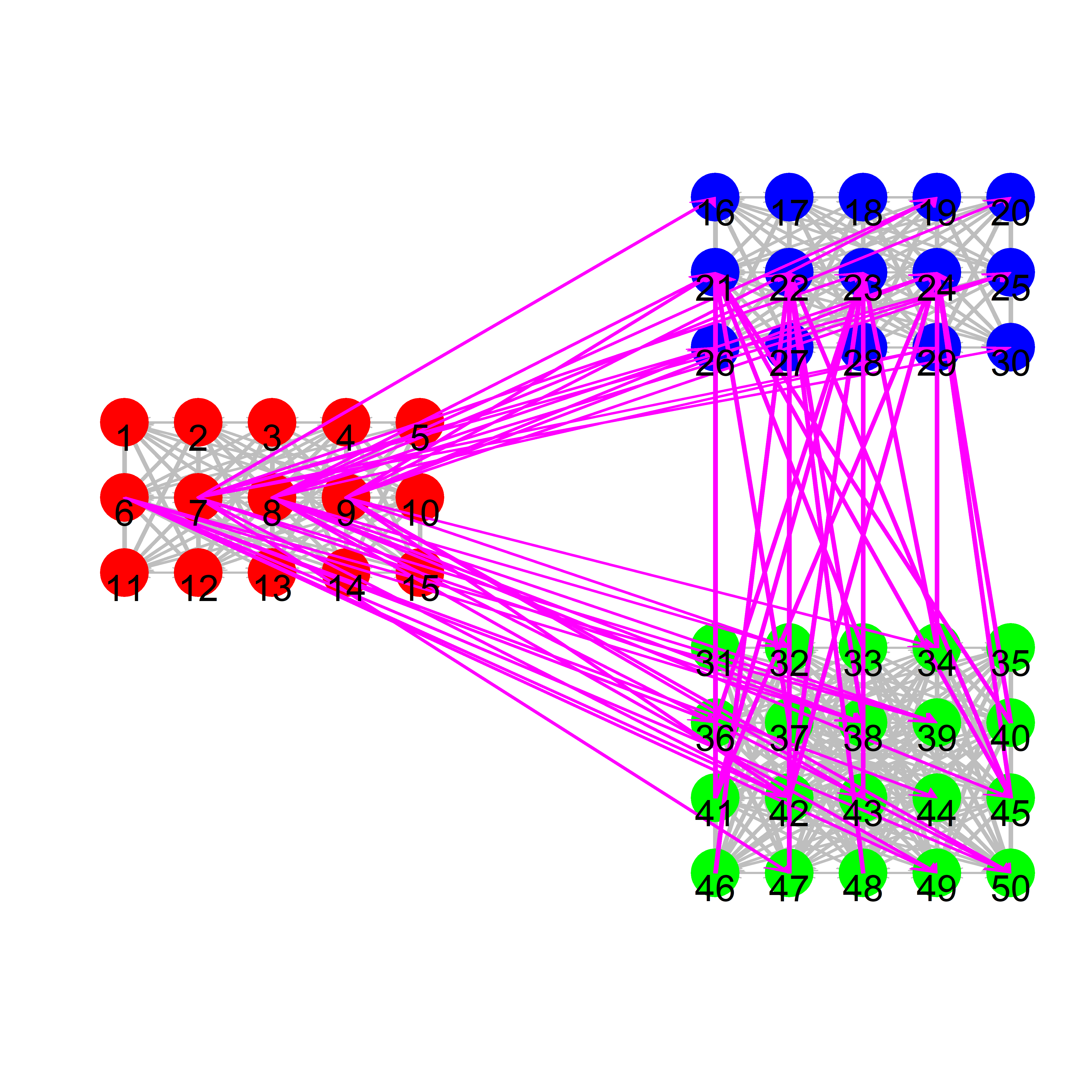}
    }
    \subfigure[\footnotesize{ROC curves}]
    {\label{fig:ROCDCM}
    \includegraphics[height=4.00cm,width=5.2cm,trim= 0mm -0.15mm 0mm 0mm,clip=TRUE]{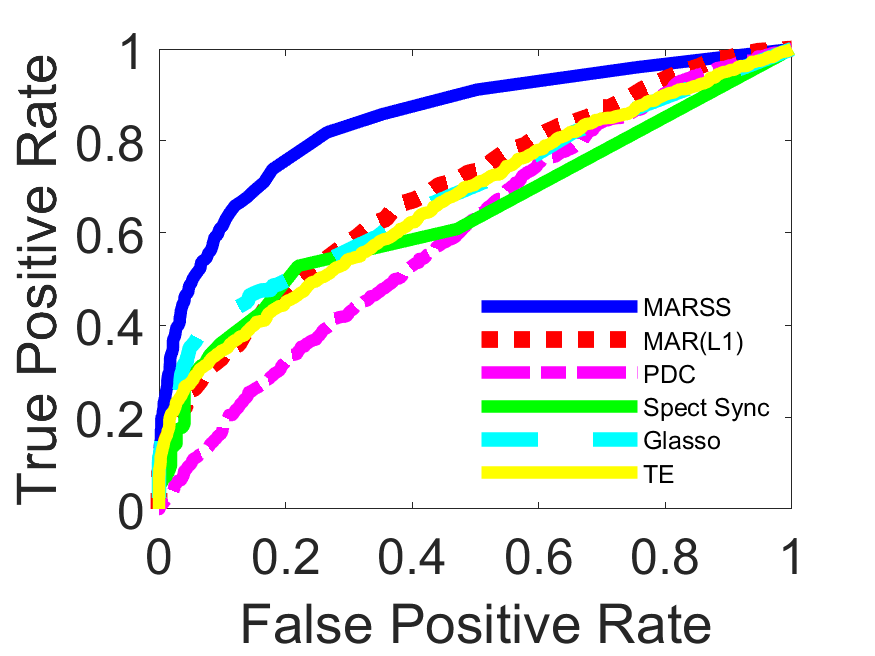}
    }
    \subfigure[\footnotesize{ROC curves}]
    {\label{fig:ROCDCMfMRI2}
    \includegraphics[height=4.00cm,width=5.2cm,trim= 0mm -0.15mm 0mm 0mm,clip=TRUE]{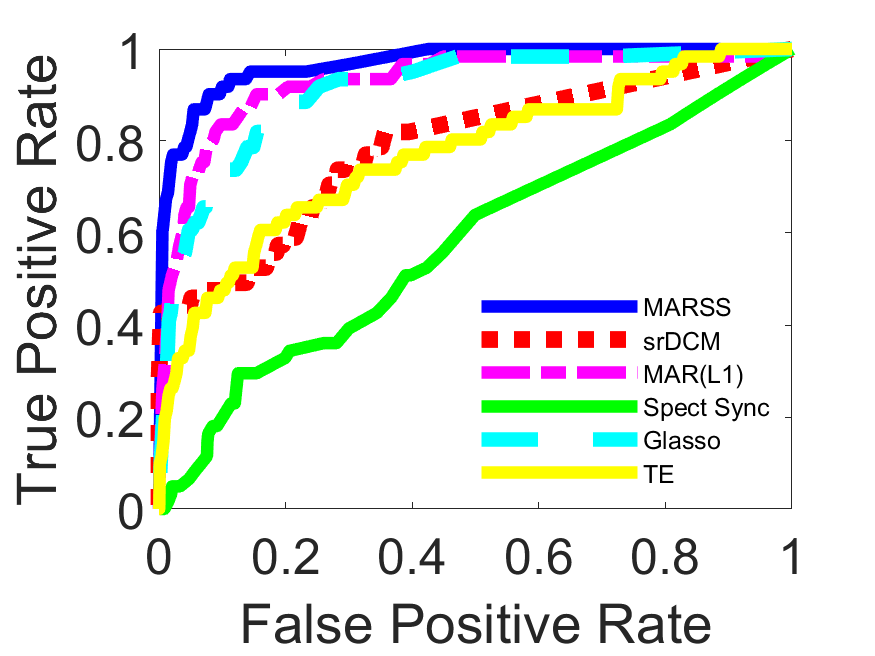}
    }\vspace{-0.4cm}
\caption{Simulation studies of data generated from DCMs. (a) The simulated network from the DCM for EEG data. (b) ROC curves
for directional connections identified by 6 network methods for the data generated in (a). (c) ROC curves for directional connections identified by 6 network methods for fMRI data simulated from a whole-brain directional network model.}
\end{figure}

 In the first simulation study, we used the neuronal state equations (which consist of many ordinary differential equations) in the DCM for EEG data (DCM-EEG) \citep{kiebel2008dynamic} to generate the state functions $\mathbf{x}(t)$, because both EEG and intracranial EEG signals are mainly from pyramidal neurons. The state equations characterized a directional network among 50 regions  with 3 clusters. Figure \ref{fig:sys5} shows the simulated network, where an edge indicates the presence of a directional connection between a pair of regions. Based on the generated $\mathbf{x}(t)$, we used the observation model \eqref{eq:obseq} to simulate $\mathbf{y}(t)$, where the data error $\epsilon_i(t)$ were spatially and temporally correlated with between-region and lag-1 temporal correlations as 0.5. As such, the data-generating model was distinct from the assumed MARSS. We applied our Bayesian method to the simulated data and obtained network edge probabilities for all the directional connections.

Given the simulated directional connections (directional edges) as the true values, we calculated false positive rates (FPR) and true positive rates (TPR) of using different thresholds for network edge probabilities to identify directional connections. For comparison, we also examined
the FPRs and TPRs of other network methods, including the multivariate autoregression with $L_1$ regularization, denoted by MAR($L_1$), partial directed coherence (PDC) \citep{baccala2001partial}, the spectrum synchronicity \citep{euan2018spectral}, transfer entropy (TE) \citep{vicente2011transfer}, and graphical lasso (Glasso) \citep{friedman2008sparse,friedman2014glasso,witten2011new}. Note that the DCM \citep{kiebel2008dynamic} could not produce the network estimate because its estimation for 50 regions is computationally infeasible.  The ROC curves of TPRs vs. FPRs for these methods are shown in Figure \ref{fig:ROCDCM}. The new MARSS outperformed other methods by achieving a much larger area under the curve.

In the second simulation study, we compared the new MARSS with existing directional network methods using fMRI data simulated from a whole-brain directional network model (DCM-fMRI) \citep{frassle2018generative}. The simulated network (shown in \citep{frassle2018generative}) had the small-world architecture of the human brain. Though the new MARSS is distinct from the DCM-fMRI, the new MARSS achieved the largest area under the ROC curve, as shown in Figure \ref{fig:ROCDCMfMRI2}. Note that the new MARSS also outperformed the sparse regression-DCM (srDCM) \cite{frassle2018generative}, which is a recent extension of the DCM-fMRI to high-dimensional directional networks.

\subsection{Analysis of Intracranial EEG Recordings}

\subsubsection{Increase in the number of directional connections of SOZ after seizure onset}
\begin{figure}
\centering
   \subfigure[Electrode Placement]
   {\label{fig:P1}
   \includegraphics[height=3.8cm,width=5.1cm,trim= 50mm 20mm 45mm 45mm,clip=TRUE]{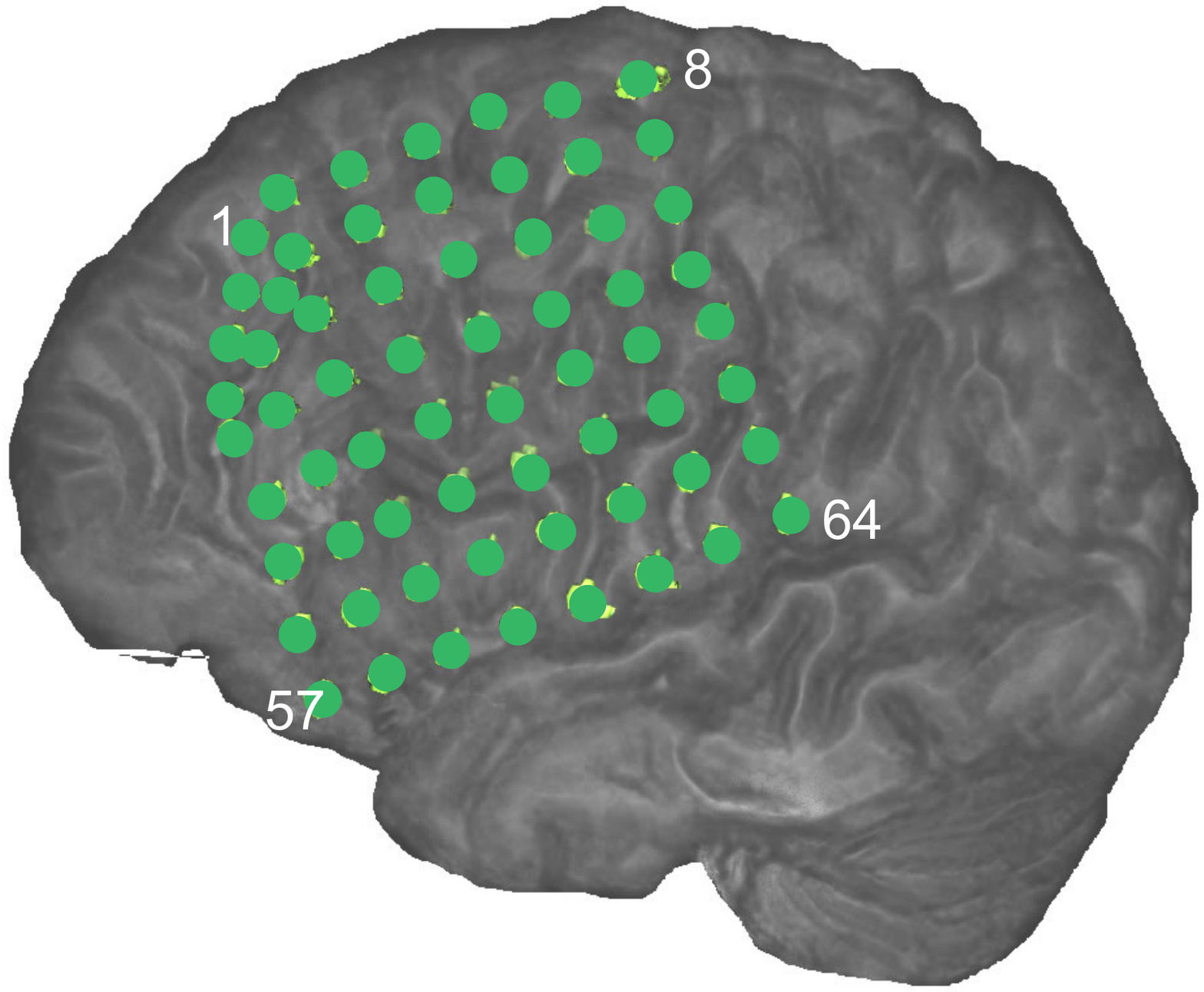}
    }
   \subfigure[275-300 seconds before onset]
   {\label{fig:275bP1}
   \includegraphics[height=3.8cm,width=5.1cm,trim= 34mm 48mm 26mm 25mm,clip=TRUE]{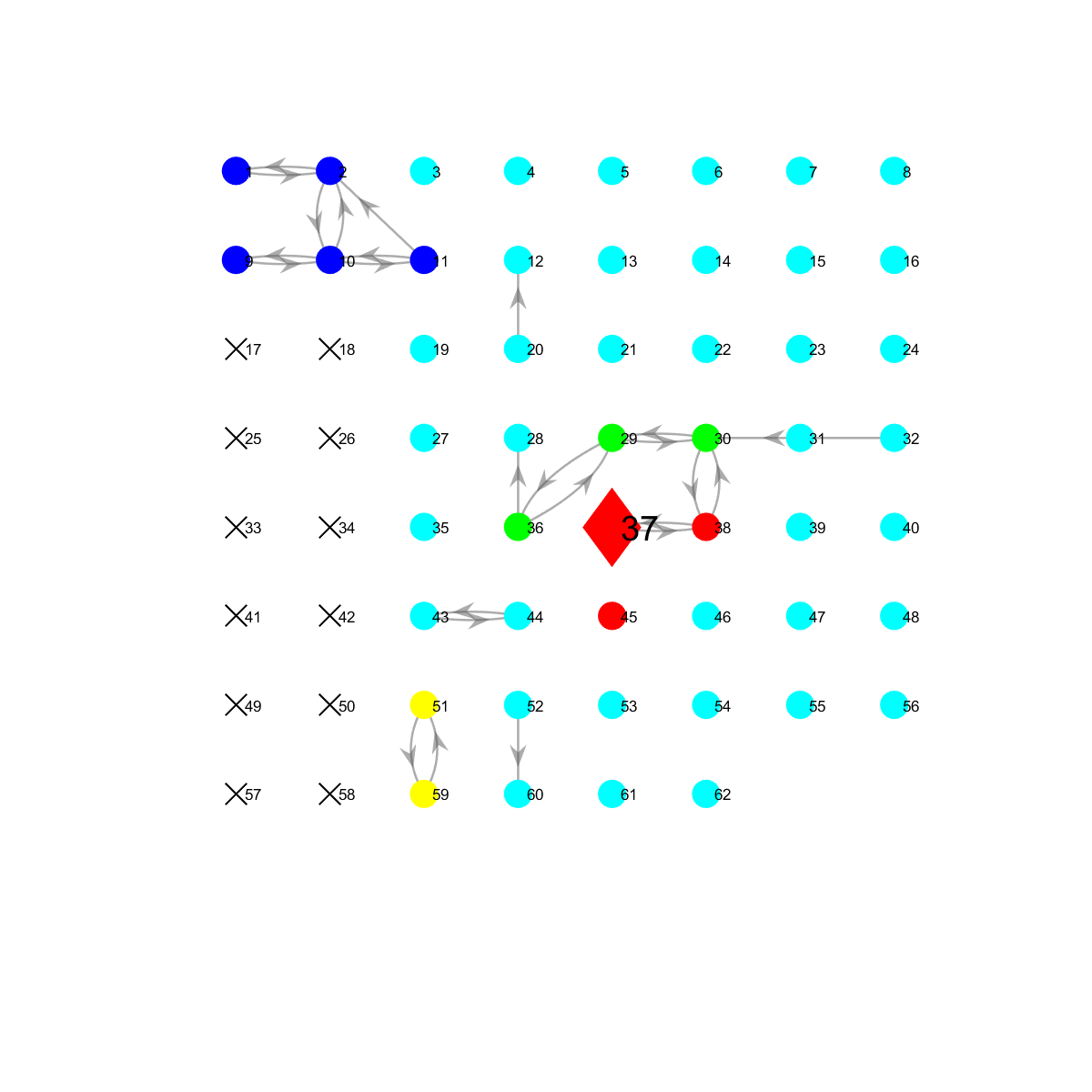}
    }
        \subfigure[0-25 seconds before onset]
   {\label{fig:0bP1}
   \includegraphics[height=3.8cm,width=5.1cm,trim= 34mm 48mm 26mm 25mm,clip=TRUE]{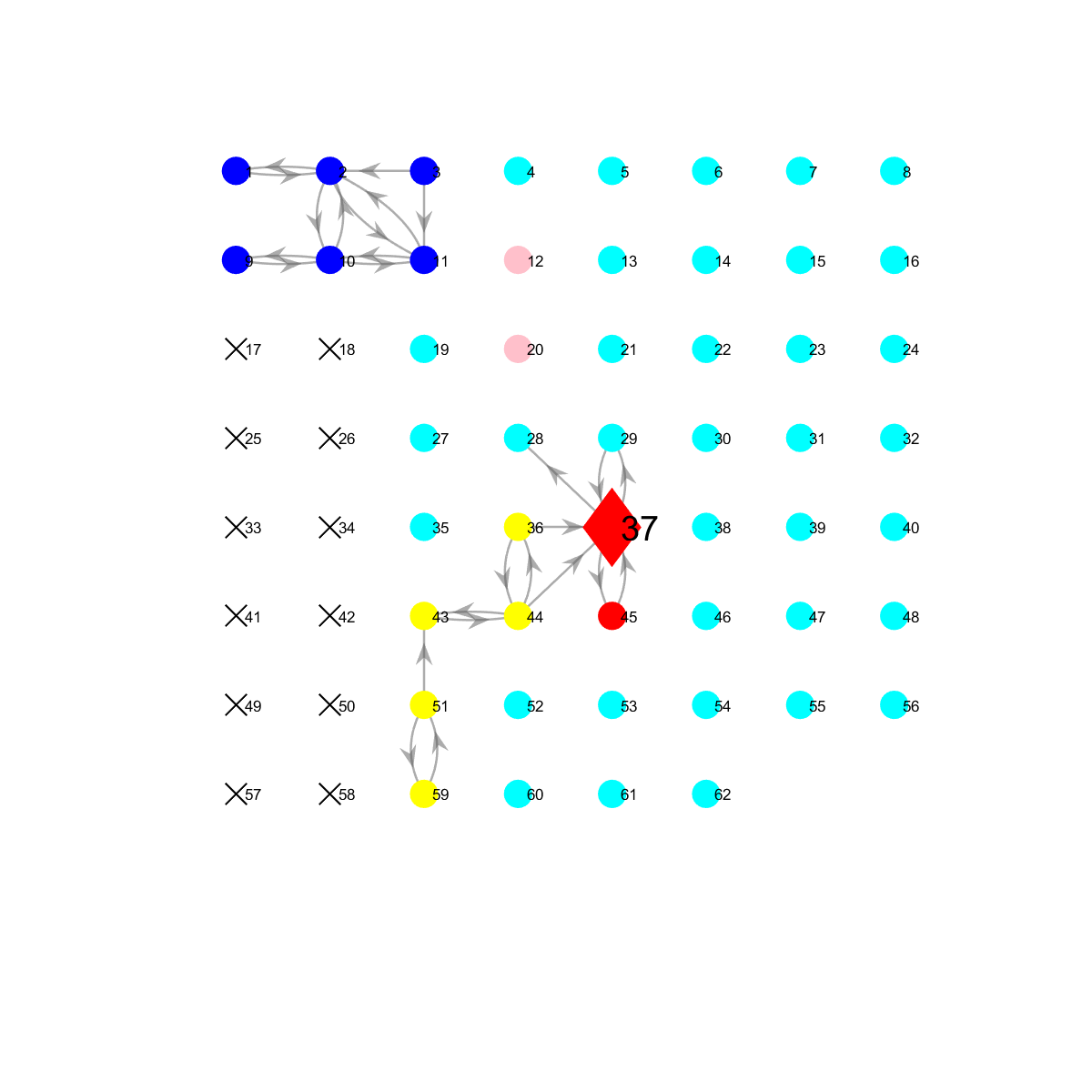}
    }\\
       \subfigure[0-25 seconds after onset]
   {\label{fig:0aP1}
   \includegraphics[height=3.8cm,width=5.1cm,trim=34mm 48mm 26mm 25mm,clip=TRUE]{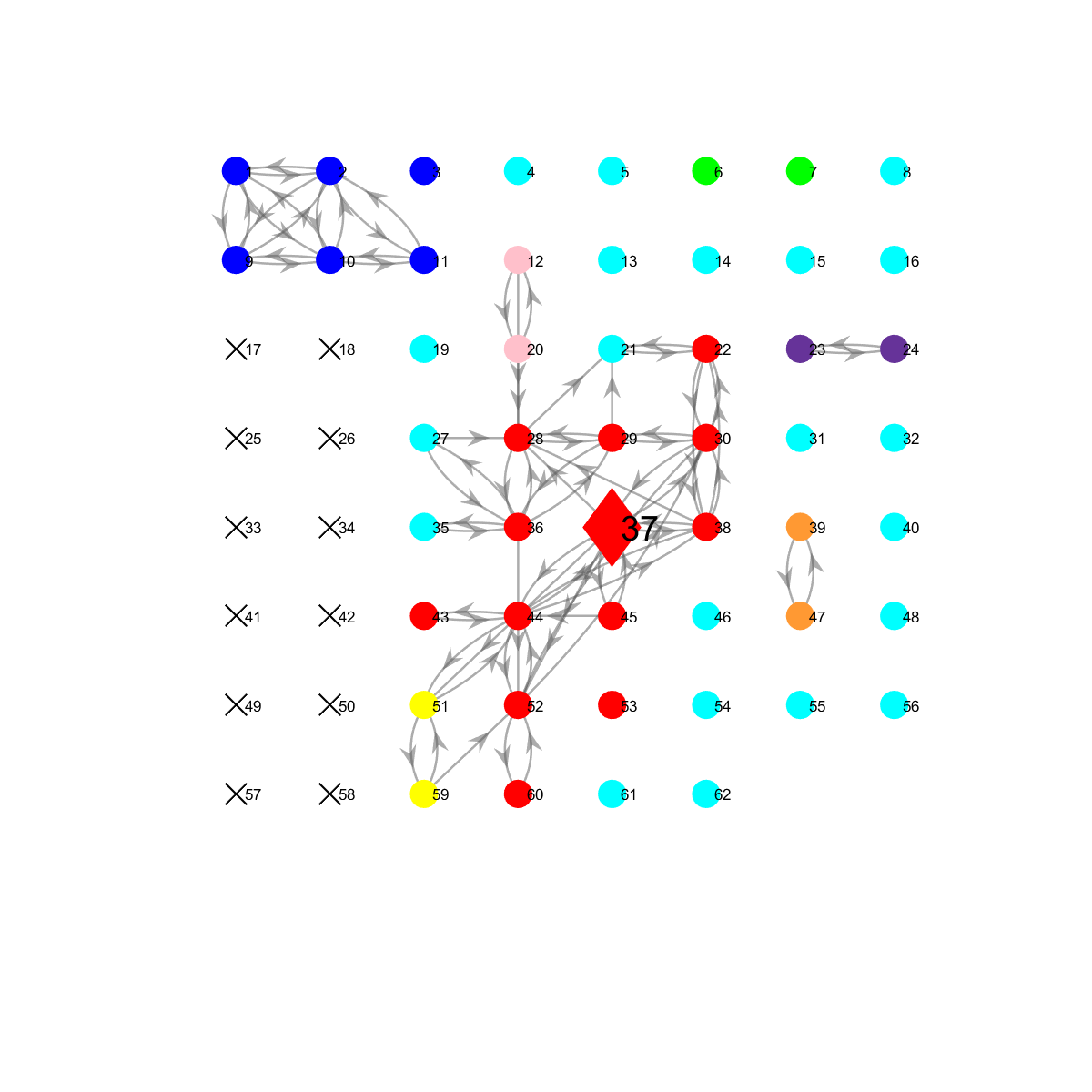}
    }
         \subfigure[75-100 seconds after onset]
   {\label{fig:25aP1}
   \includegraphics[height=3.8cm,width=5.1cm,trim= 34mm 48mm 26mm 25mm,clip=TRUE]{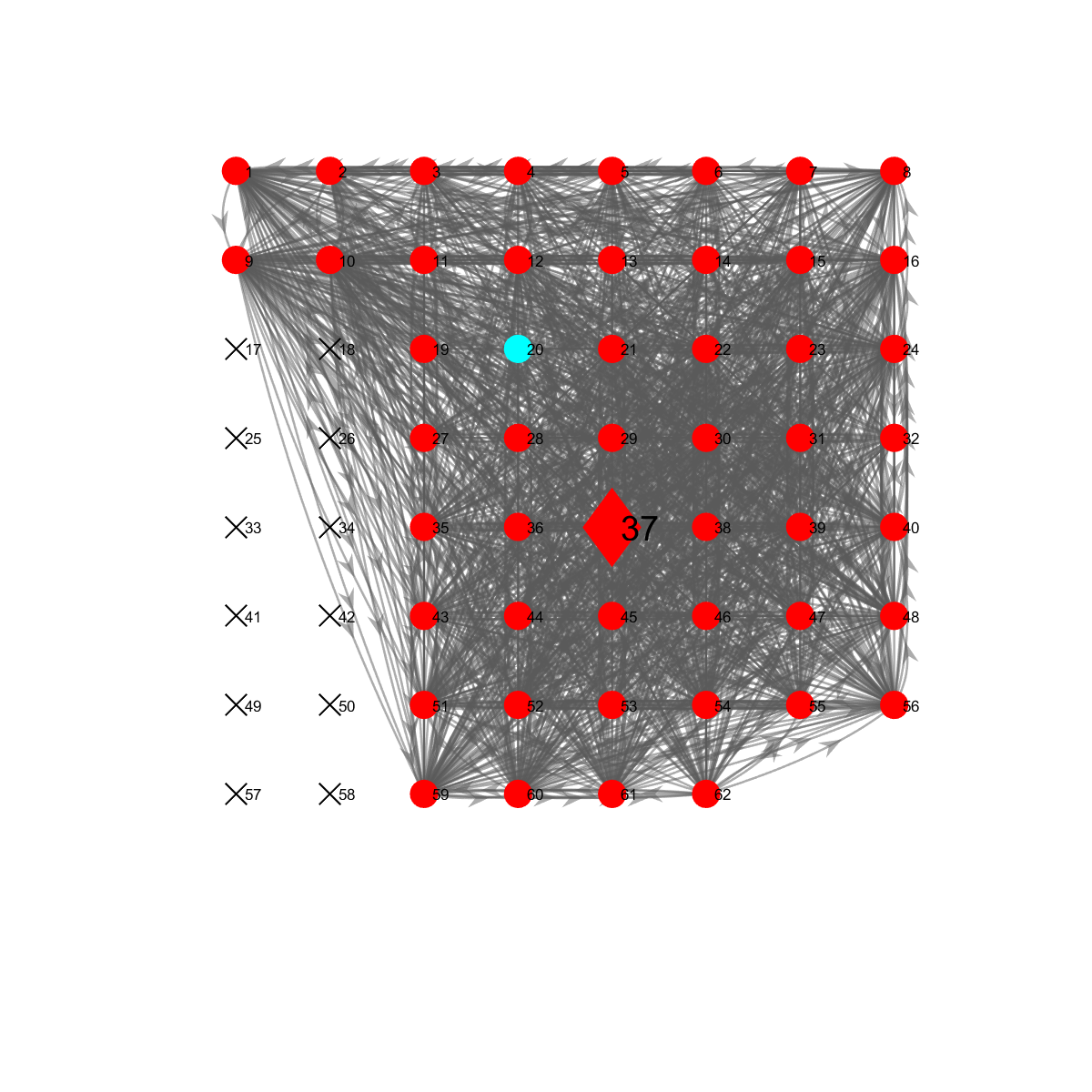}
    }
              \subfigure[200-225 seconds after onset]
   {\label{fig:225aP1}
   \includegraphics[height=3.8cm,width=5.1cm,trim= 34mm 48mm 26mm 25mm,clip=TRUE]{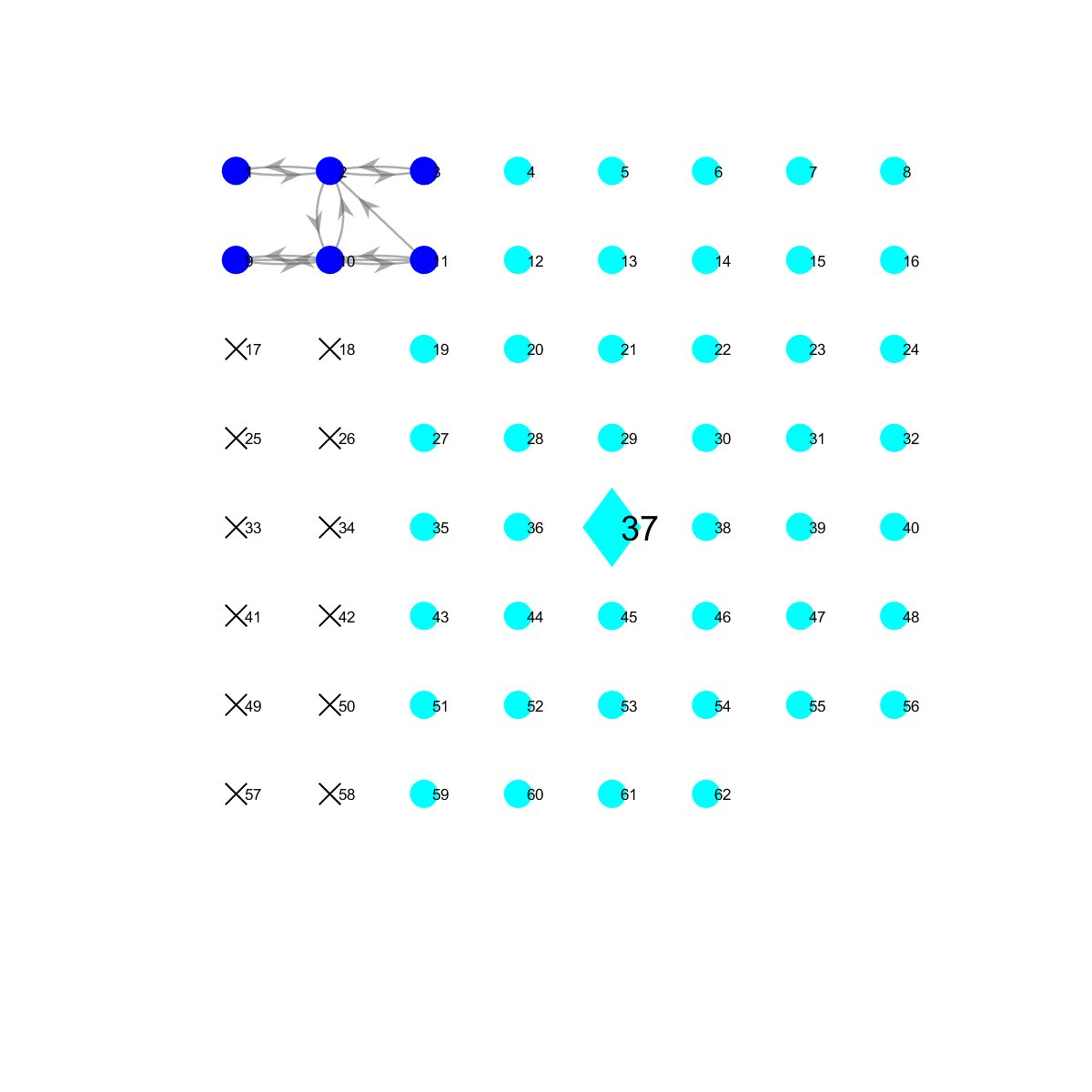}
    }
        \caption{\label{fig:Sub1} Patient 1's directional brain networks identified by the new MARSS. (a) The intracranial EEG electrode placement on the left hemisphere of Patient 1. (b)-(f) The identified directional brain networks from 300 seconds before to 225 seconds after seizure onset. The diamond at electrode G37 is the SOZ identified by expert interpretation of EEG data. Nodes in light blue are the regions that did not belong to any clusters. Nodes in the same other colors (either dark blue, green, pink, red, purple, brown or yellow) denote different identified clusters of regions. All nodes in red color belong to the SOZ cluster. Grey arrows indicate the identified directional connections between regions. Anterioinferior electrodes preceded by an ``X" were resected in a previous epilepsy surgery. The evolution of the brain network from 300 seconds before to 300 seconds after seizure onset is shown as a video in the supplementary files.}
\end{figure}

For all patients analyzed, we found that their interictal directional brain networks were stable up to the seizure onset time. Moreover, directional connections were sparse in these interictal brain networks. Specifically, before seizure onset (Figures \ref{fig:275bP1} and \ref{fig:0bP1}), the (clinically determined) SOZ was connected to just a few regions, most of which were immediately adjacent to the SOZ.  However, immediately after seizure onset (in the window of 0-25 seconds after seizure onset), the number of directional connections significantly increased (Figure \ref{fig:0aP1}) with P-values $\leq$0.03 for all patients (see details of hypothesis testing procedures in Appendix). The majority of these new directional connections  occurred in the area of the SOZ and its adjacent regions. The number of directional connections continued to grow as seizure progressed (Figure \ref{fig:25aP1}).  The cessation of seizure activity was marked by a return to the number of directional connections in interictal phases (Figure \ref{fig:225aP1}). Overall, the new MARSS revealed the evolution of the directional brain network from a normal to abnormal epileptic state and then back to the normal state.

\subsubsection{Comparison with other network approaches} We compared directional connections identified by the new MARSS with those identified by three common network methods, including: i) short-time direct transfer function (SdDTF) \citep{korzeniewska2014ictal}, ii) partial directed coherence (PDC) \citep{baccala2001partial}, and iii) cross-coherence (CC) \citep{varela2001brainweb}. SdDTF and PDC measure directional connectivity. CC measures function connectivity without directionality information. To assess the performance of the new MARSS and these three methods in detecting brain network changes at the seizure onset time, we compared the number of identified connections (with or without direction) of the SOZ across three 25-second windows: 275-300 seconds before seizure onset (\emph{baseline}), 0-25 seconds before seizure onset (\emph{immediate preictal}), and 0-25 seconds after seizure onset (\emph{immediate postictal}).

As shown in Table \ref{table:}, the new MARSS consistently detected increases in the number of directional connections of the SOZ in the immediate postictal window, compared to the preictal windows, for all 6 patients analyzed. In contrast, none of the three competing methods (SdDTF, PDC, or CC) consistently detected such increases. SdDTF detected increases in the number of directional connections in the immediate postictal window, compared to the baseline and immediate preictal  windows, for Patients 1, 3, 5, and 6, but failed to show similar increases in the directional network for Patients 2 and 4. PDC failed to detect more directional connections in the immediate postictal window versus the other two interictal windows for Patients 4 and 6. CC detected zero connections in all the three windows for Patients 1, 3, and 4. Overall, the new MARSS outperformed the other three methods by consistently detecting changes in the directional brain networks around the SOZs at seizure onset.

\begin{table*}
\begin{center}
\caption{Comparison of Different Network Methods \label{table:}}
\begin{tabular}{|c|c|cccccc|}
\hline
Methods               &Windows&\multicolumn{6}{c|}{Patients}   \\
                      &         &1   & 2  &3  &4  &5 &6        \\ \hline

\multirow{6}{*}{MARSS}&Baseline &2   & 10 &3  &0  &8 &16   \\       \cdashline{2-8}

                      &Immediate&    &    &   &   &  & \\
                      &Preictal & 7  &9   &0  &0 &14 & 45   \\  \cdashline{2-8}
                      &Immediate&    &    &   &  &   & \\

                      &Postictal&11 &15 &7  &6  &63 &220  \\
\hline
\multirow{6}{*}{PDC} &Baseline &6   & 0 &3  &1  &15 &29  \\       \cdashline{2-8}

                     &Immediate&    &    &   &   &  & \\
                     &Preictal & 6  &0   &4  &0 &10 & 18   \\  \cdashline{2-8}
                     &Immediate&    &    &   &  &   & \\

                     &Postictal&8 &1  &7  &1  &29 &23  \\
\hline
\multirow{6}{*}{SdDTF}&Baseline &15   & 0 &0  &2  &20 &75 \\       \cdashline{2-8}

                     &Immediate&    &    &   &   &  & \\
                     &Preictal & 7  &2   &1  &1 &35 & 12   \\  \cdashline{2-8}
                     &Immediate&    &    &   &  &   & \\

                     &Postictal&19 &2  &6  &1  &53 &87  \\
\hline
\multirow{6}{*}{CC}  &Baseline &0   & 4 &0  &0  &14 &11 \\       \cdashline{2-8}

                     &Immediate&    &    &   &   &  & \\
                     &Preictal & 0  &6   &0  &0 &30 & 23   \\  \cdashline{2-8}
                     &Immediate&    &    &   &  &   & \\

                     &Postictal&0 &9  &0 &0  &152 &49  \\
\hline
\end{tabular}\end{center}
\footnotesize{The numbers of directional connections of the SOZ identified by the new MARSS versus 3 other methods, including: i) short-time direct transfer function (SdDTF), ii) partial directed coherence (PDC), and iii) cross-coherence (CC) for 6 patients in the baseline (275-300 seconds before seizure onset), immediate preictal (0-25 seconds before seizure onset), and immediate postictal (0-25 seconds after seizure onset) windows.}
\end{table*}

\subsubsection{Increase in the number of regions in the SOZ cluster} The new MARSS enables identifying clusters of a large number of regions. In this study, the SOZ cluster is of particular interest because it is a group of regions that are most affected by the activity in the SOZ. With the new MARSS, we identified the SOZ cluster and revealed its changes from interictal to ictal phases to demonstrate the effect of seizure propagation on changing the brain network structure.

For all patients analyzed, their SOZ cluster sizes (the number of regions
contained in the identified SOZ cluster) remained stable over time in interictal phases. As shown in Figures \ref{fig:275bP1} and \ref{fig:0bP1}, the SOZ cluster often included only the SOZ and a few immediately adjacent regions, indicating that very few regions were affected by activity in the SOZ before seizure onset. However, immediately after seizure onset, the SOZ cluster expanded significantly (P-values $\leq$0.02 for all patients) to include more nearby regions (Figures \ref{fig:0aP1}). The SOZ cluster size continued to increase as seizure progressed (Figure \ref{fig:25aP1}). With seizure cessation, similar to the changes in the number of directional connections, the size of the SOZ cluster returned to the level in interictal phases.

\begin{figure}
\centering
 \subfigure[Intracranial EEG of Patient 2]
   {\label{fig:P2EEG}
   \includegraphics[height=6.8cm,width=15.0cm,trim=29mm 0mm 30mm 12mm,clip=TRUE]{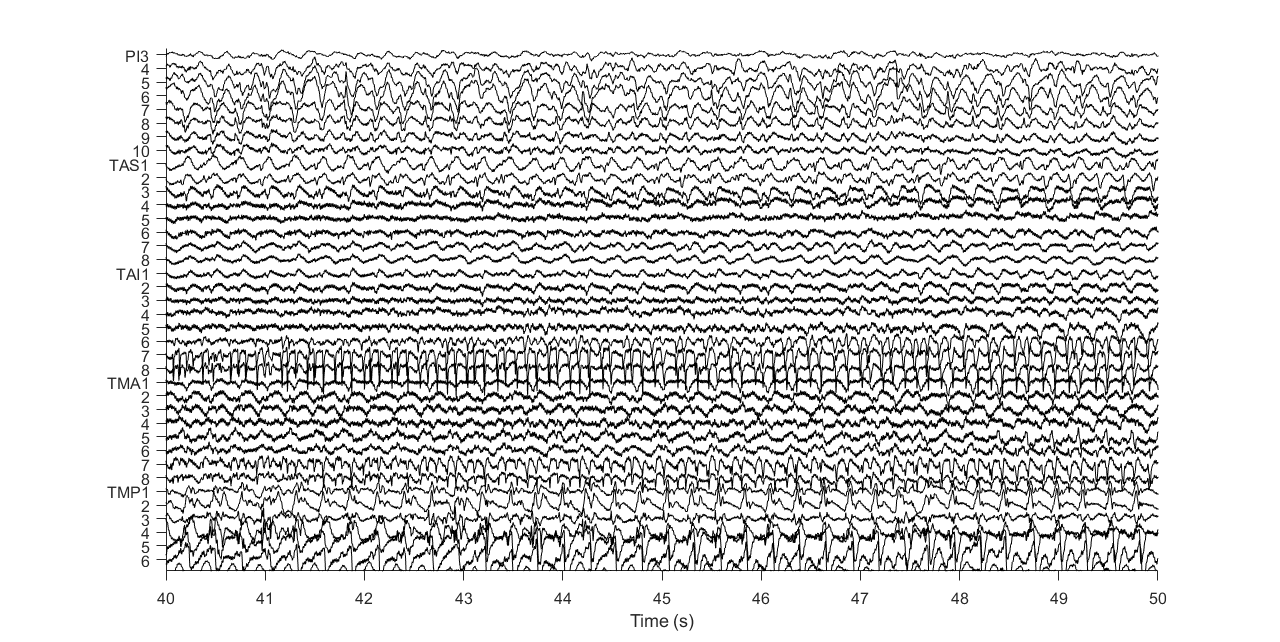}
    }
           \subfigure[25-50 seconds before onset]
   {\label{fig:25bP2}
   \includegraphics[height=5.4cm,width=7.5cm,trim= 0mm 0mm 0mm 0mm,clip=TRUE]{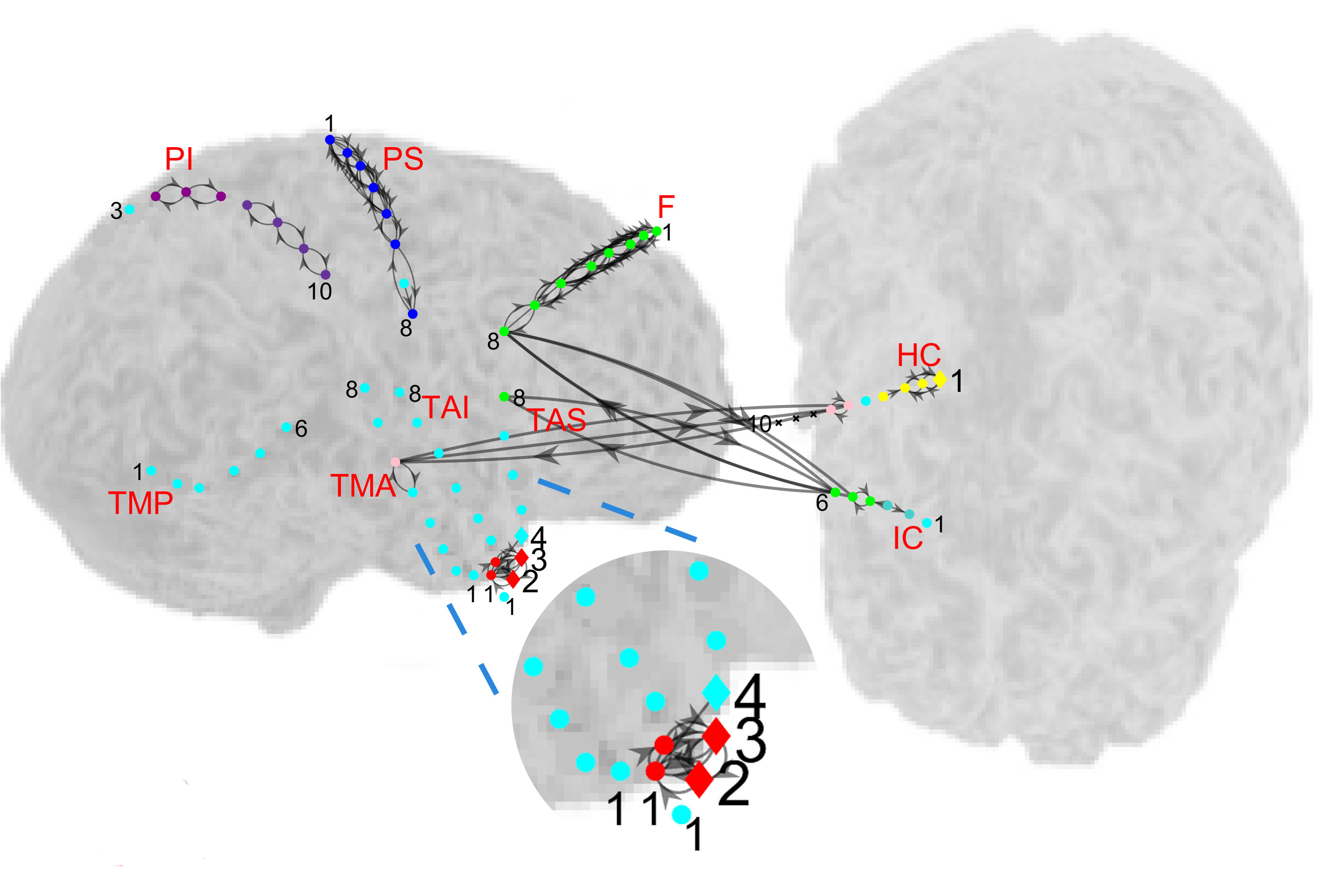}
  }
   \subfigure[0-25 seconds before onset]
   {\label{fig:0bP2}
   \includegraphics[height=5.4cm,width=7.5cm,trim= 0mm 0mm 0mm 0mm,clip=TRUE]{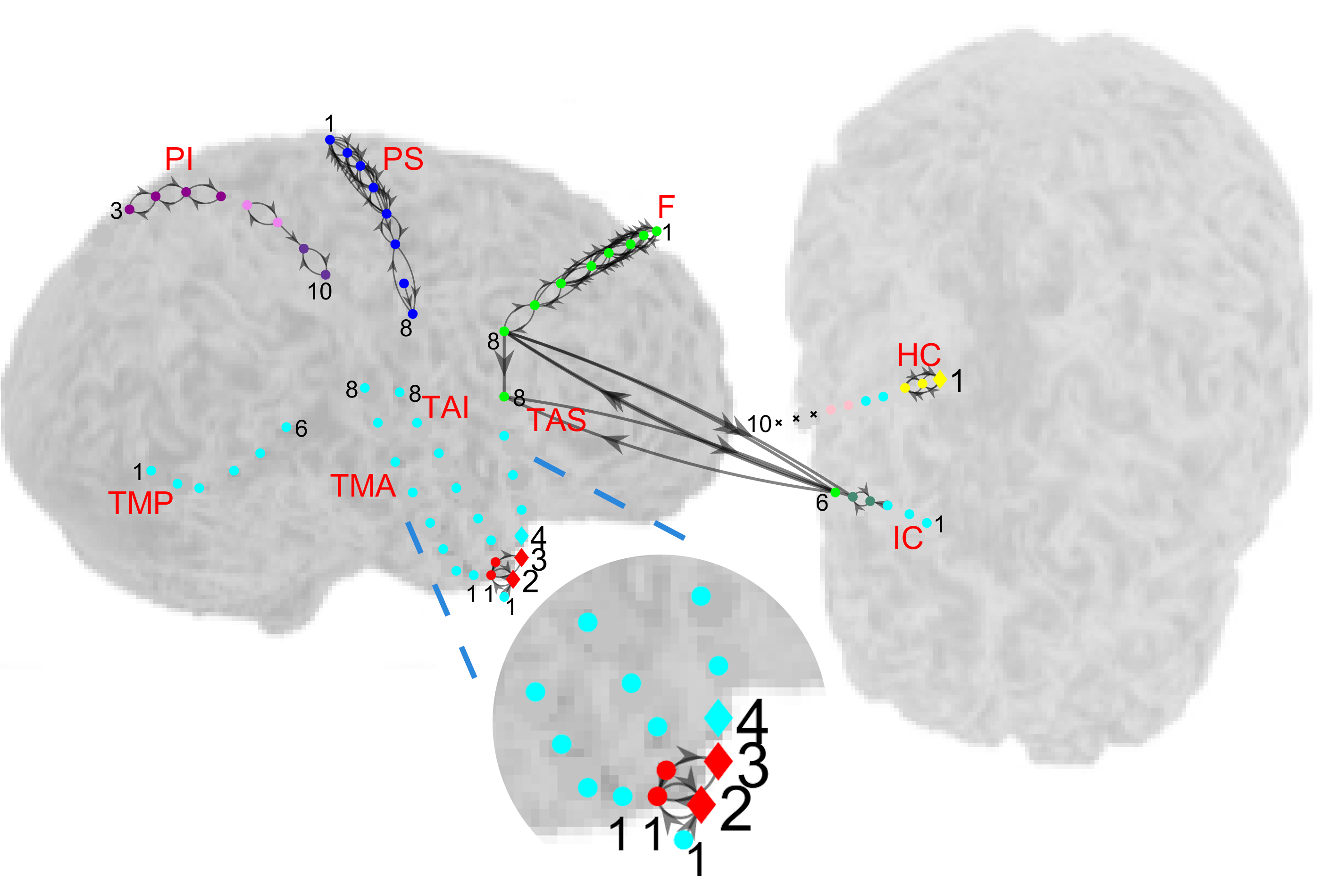}
  }\\ \vspace{-0.2cm}
          \subfigure[0-25 seconds after onset]
   {\label{fig:0aP2}
    \includegraphics[height=5.4cm,width=7.5cm,trim= 0mm 0mm 0mm 0mm,clip=TRUE]{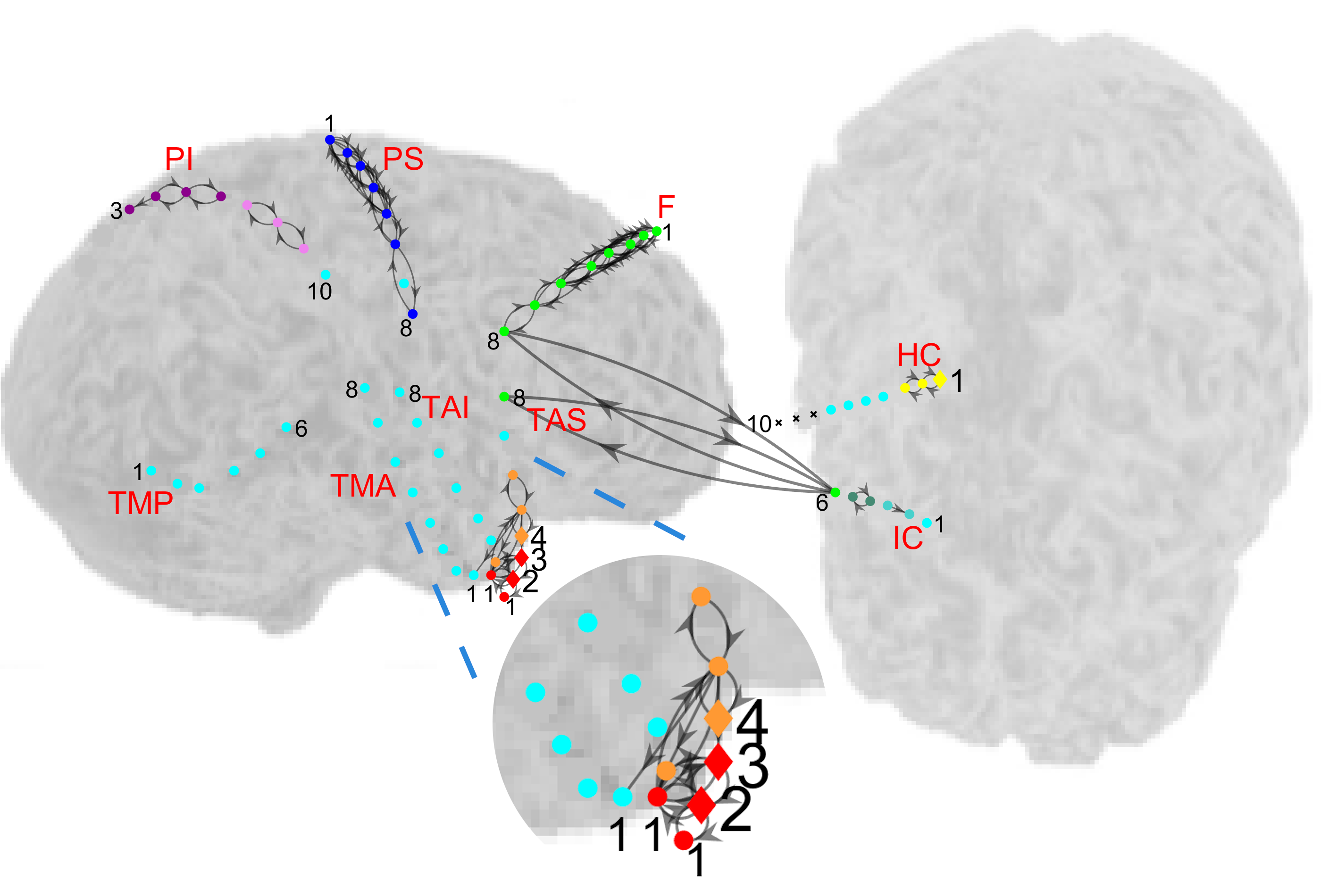}
    }
         \subfigure[25-50 seconds after onset]
   {\label{fig:25aP2}
    \includegraphics[height=5.4cm,width=7.5cm,trim= 0mm 0mm 0mm 0mm,clip=TRUE]{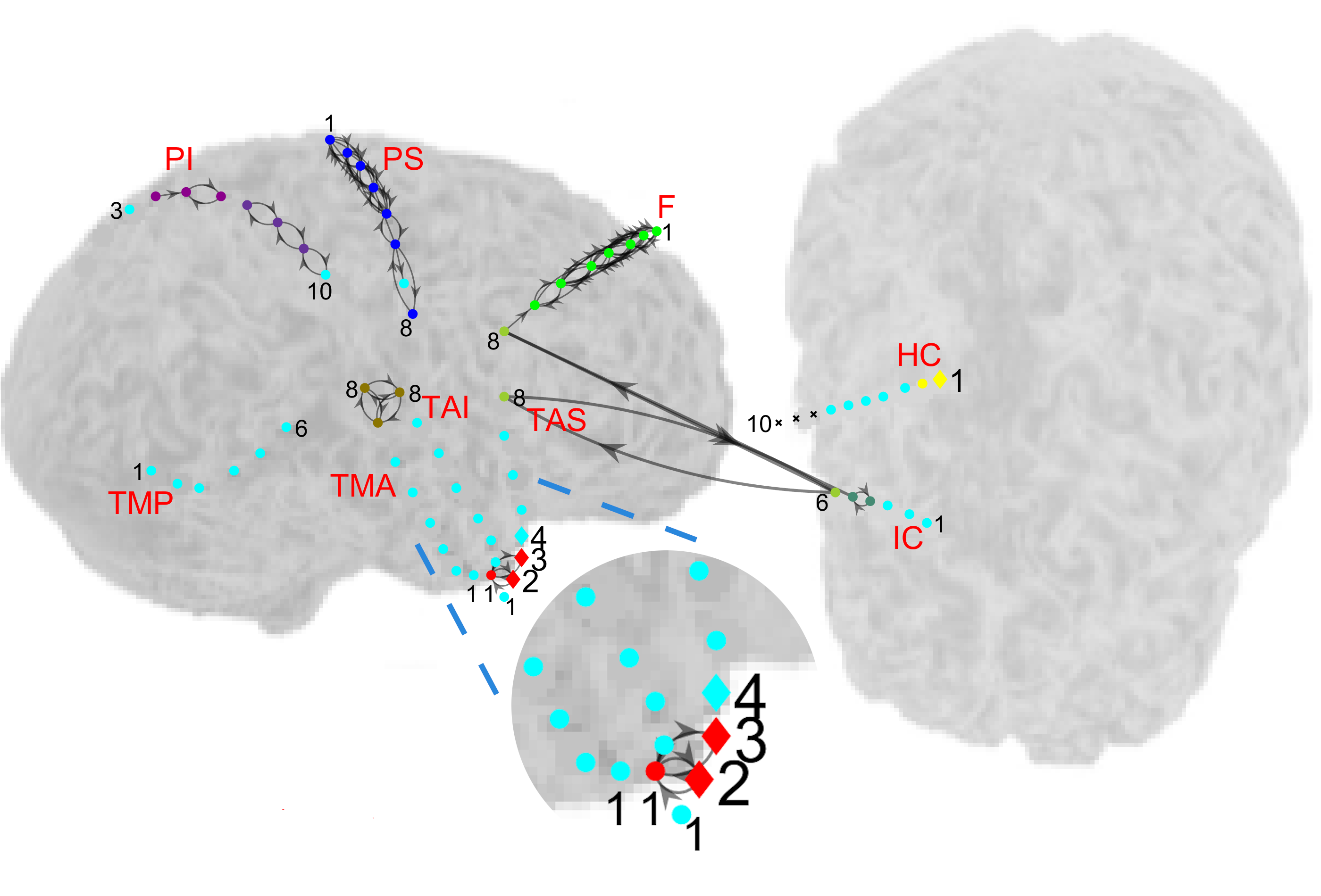}
    }\vspace{-0.4cm}
\caption{\label{fig:Sub2} Patient 2's directional brain networks identified by the new MARSS. (a) A segment of intracranial EEG recordings of Patient 2. (b)-(e) The identified directional brain networks
from 50 seconds before to 50 seconds after seizure onset. The diamonds are the SOZ identified by expert interpretation of EEG data. Nodes in light blue are the regions that did not belong to any clusters. Nodes in the same other colors denote different identified clusters of regions. Arrows indicate the identified directional connections between regions. The evolution of the brain network is shown as a video in the supplementary files.}
\end{figure}

\subsubsection{Changes in directional connections of non-SOZ regions} Most previous studies focused on the changes in the directional brain network around the SOZ only \citep{Jirsa2014}. With the new MARSS, we distinguished among the SOZ, non-SOZ regions in the SOZ cluster, and other non-SOZ outside the SOZ cluster. We found that the latter two types of non-SOZ regions experienced different changes in directional connections during the transition from interictal to ictal phases. Regions in the SOZ cluster (regardless of time window) all had substantial increases in the number of directional connections in ictal phases compared to interictal phases. For example, for Patient 1 (electrode placement is shown in Figure \ref{fig:P1}), all the regions in the SOZ cluster (electrodes in red) had significant increases in the number of directional connections during 75-100 seconds after seizure onset (Figure \ref{fig:25aP1}), compared to their numbers of directional connections in interictal windows.

On the other hand, many regions outside the SOZ cluster did not exhibit increases in directional connections though these regions displayed ictal activity based on visual interpretation of EEG. For example, for Patient 2 (electrode placement is shown in Figure \ref{fig:25bP2}), regions outside the SOZ cluster, such as TMA3-TMA5 and TMP3-TMP6, had no direct connections with the SOZ in both interictal and ictal phases. These regions had no changes in directional connections from interictal to ictal phases (Figures \ref{fig:25bP2}-\ref{fig:25aP2}), despite that their intracranial EEG data signaled ictal activity during seizure propagation (Figure \ref{fig:P2EEG}). This result indicates that not all regions showing epileptic activity are
actually involved in seizure propagation or directly influenced
during seizure propagation.

\subsubsection{Directional connectivity analysis for SOZ localization}
\begin{figure*}
\centering
   \subfigure[Electrode Placement]
   {\label{fig:P3}
   \includegraphics[height=4.8cm,width=5.0cm,trim= 0mm 0mm 0mm 2mm,clip=TRUE]{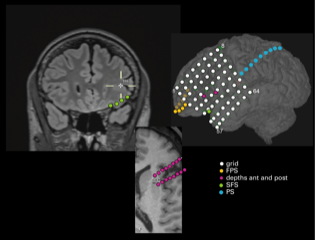}
   }
   \subfigure[Connectivity Changes $t\in (-275,275)$]
   {\label{fig:Sub3ADC}
   \includegraphics[height=4.8cm,width=5.0cm,trim= 0mm 5mm 10mm 17mm,clip=TRUE]{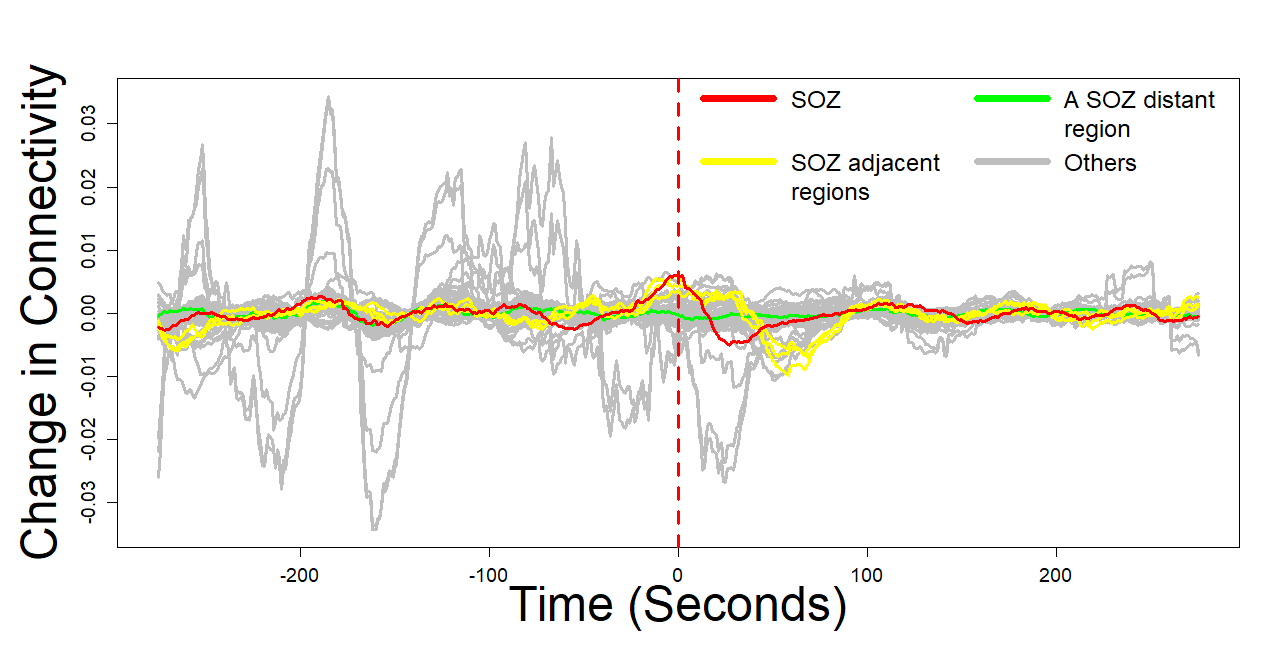}
   }
   \subfigure[Connectivity Changes $t\in(-50,50)$]
   {\label{fig:Sub3ADCTrim}
   \includegraphics[height=4.8cm,width=5.0cm,trim= 0mm 5mm 10mm 17mm,clip=TRUE]{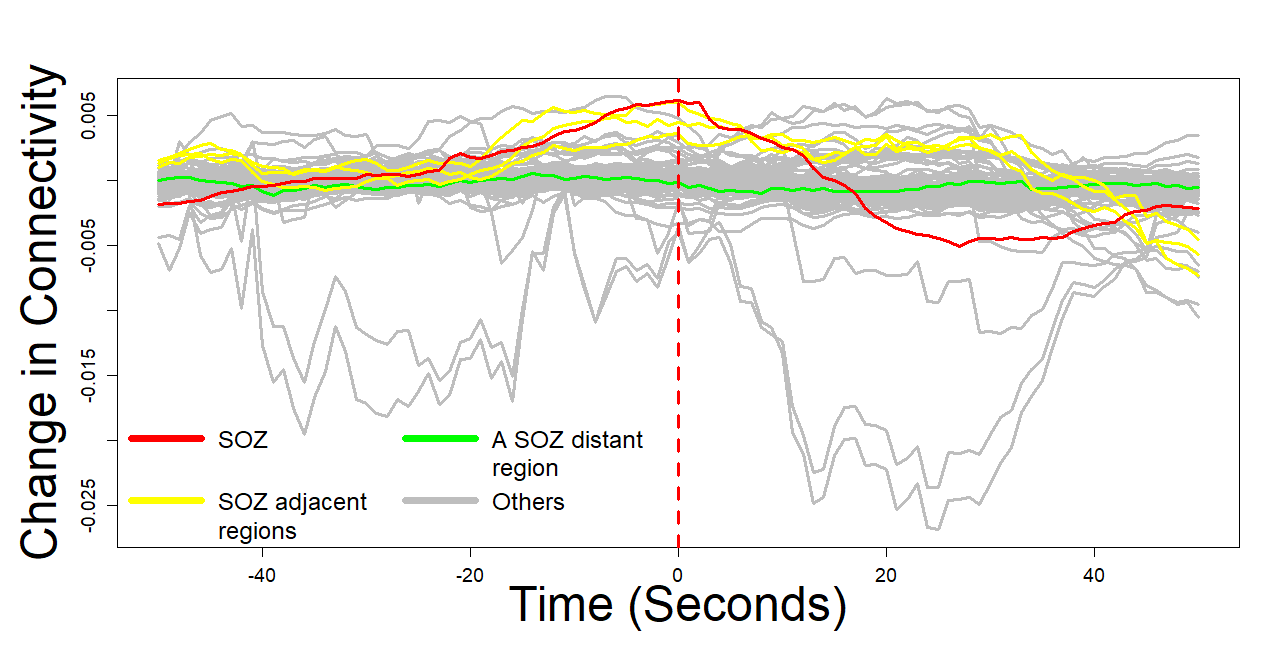}\vspace{-0.5cm}
   }
   \caption{The directional connectivity changes over time of all regions for Patient 3. (a) Electrode placement of Patient 3. (b) The time series of directional connectivity changes, $DC^t_j$, of all regions from 275 seconds before to 275 seconds after seizure onset. $t=0$ is the seizure onset time. (c) The time series of  directional connectivity changes of all regions from 50 seconds before to 50 seconds after seizure onset. The red curve is the directional connectivity changes of the SOZ.}
\end{figure*}

The above results show the identified changes with respect to the overall directional network structure from interictal to ictal phases. We next show the change in directional connectivity for individual regions. Figure \ref{fig:Sub3ADC} presents directional connectivity changes ($DC^t_j$ defined in Methods and Materials) of all individual regions for Patient 3 from 275 seconds before to 275 seconds after seizure onset, with Figure \ref{fig:Sub3ADCTrim} focused on the shorter time interval around the seizure onset time ($\pm$50 seconds around seizure onset). We found that at $t=0$ when seizures started, the SOZ and its a few adjacent regions experienced the greatest directional connectivity changes. Based on this unique directional connectivity property of the SOZ, we developed a SOZ localization method (Methods and Materials). Using this method, we selected G30 and its three adjacent regions (G22-G24) to be candidate regions for the SOZ. Since G30 is the true SOZ, we regard this selection accurately identified the same region as verified by expert visual interpretation of EEG for Patient 3.

We assessed the effectiveness of our method in detecting the unique directional connectivity property of the SOZ by evaluating the accuracy of our method in localizing SOZs for all 6 patients. We used clinically localized SOZs as the given truth. We deemed our method to successfully detect one SOZ site (i.e., a true positive) if at least one of the regions selected by the method was within one electrode of the SOZ site. A false positive was scored when a region selected by the method was beyond one electrode of any SOZ site. Our localization method achieved 100\% true positive rates (TPR) for all 6 patients (Table \ref{table:SOZLocal}) despite their different SOZ locations and different numbers of regions inside the SOZ areas. The false positive rates (FPR) were zero for 3 patients and 2-3\% for the other 3 patients. Note that Patient 6 had multiple SOZ sites, which are difficult to accurately localize by using existing quantitative methods. In contrast, our method successfully detected all the SOZ sites for this patient.

\begin{table}
\caption{SOZ Localization \label{table:SOZLocal}  }
\centering 
\begin{tabular}{c c  c c c c c} 
\hline\hline
Patient&Clinically      &Other       & Number of         & Selected            &TPR      &FPR\\
Number &Diagnosed       &Resected    & Analyzed          &  Regions             &         &\\
       &SOZ             & Regions    &  Regions          &                      &          &  \\
\hline
1     &G37              &N/A             &50           &G28, G38,               &  100\% &2\% \\
      &                  &              &             &G30, \textbf{G37},       &        &\\
      &                   &            &              &G52                       &        &\\  \hdashline
2  &TAS2-4, HC1$^\star$ &TAS1, TAS5        &67           &\textbf{TAS2, TAS5}   &100\%   & 0\% \\\hdashline
3  &G30                 &N/A                &95           &\textbf{G30}, G23,      &100\%   &2\%\\
   &                   &                    &             &  G22, G24            &           &\\      \hdashline
4     &G17              &G16, G19          &20           &\textbf{G16, G17}           &100\%   &0\%\\  \hdashline
5  &AS3-6,            &MM3-6             &38           &\textbf{MM3, MM4,}       &100\%   &0\%\\
    & MS4-6           &                  &             & \textbf{ MS5, AS5}      &         &\\ \hdashline

6  &RTM2-4,          &  N/A             &128          &RTP1, \textbf{RTM4},&100\% & 3\% \\
  & ROD1-2,          &                  &             &    RTM5, RTP2       &       &\\
  &RTA1-3            &                  &             & RTM6, \textbf{ROD2},&      &\\
      &                 &               &             &  RTM8, \textbf{RTA1}&      &\\
      \hline
\end{tabular}\\
\footnotesize{The true positive rates (TPR) and false positive rates (FPR) of the proposed SOZ localization method in comparison to the SOZs
determined by the clinical practice. The correctly identified SOZ regions are marked in bold. $^\star$ For Patient 2,
HC1 and TAS2 were spatially close to each other and deemed to be in one SOZ site.}
\end{table}

\section{Discussion}\label{sec:Discussion}
Epilepsy is a directional network disorder \citep{englot2016regional,kramer2012epilepsy}. Existing studies of epileptic networks have mainly focused on functional connectivity without directionality information \citep{khambhati2015dynamic,stacey2019emerging} or on low-dimensional directional networks of only a few brain regions around the SOZ \citep{korzeniewska2014ictal}. In our study, we used the new MARSS to characterize the SOZ, its adjacent regions, and many more distant non-SOZ regions as one integrated high-dimensional directional network system. With the new MARSS, we identified not only local changes of the directional brain network in the SOZ's neighboring area from interictal to ictal phases, but also changes of the directional brain network at many non-SOZ regions that were spatially distant from the SOZ. Specifically, the new MARSS identified increases in the number of directional connections of the SOZ after seizure onset, along with the expansion of the SOZ cluster. These network results with respect to the SOZ were consistent with known properties of seizure propagation \citep{alarcon2012introduction,englot2016regional}. More importantly, our approach uncovered two types of regions that differed in their changes in directional connections during the transition from interictal to ictal phases: 1) Regions within SOZ clusters demonstrated substantial increases in directional connections after seizure onset, while 2) many regions outside SOZ clusters demonstrated no changes in directional connections during seizure propagation despite visual evidence of these regions' ictal activity. Earlier studies also suggested that connectivity properties of non-SOZ regions vary by their distances to the SOZ \citep{englot2015global,zaveri2009localization}. However, these studies focused on functional connectivity. The high-dimensional directional network results by the new MARSS better displayed the difference between regions that showed visual ictal activity of ``bystander" regions versus those involved in ictal propagation and thus, shed new light on the brain's normal and abnormal network mechanisms.

An important and unique feature of the new MARSS is to simultaneously identify clusters and directional connections. Both identifying connections \citep{Friston11} and identifying clusters \citep{sporns2016modular,sporns2007identification} in brain networks are of great interest in computational neuroscience, as they are critical to understanding the brain mechanism. However, identifying clusters and identifying connections are usually performed separately with different approaches, resulting in two errors in the ensuing estimated networks. The new MARSS fills this gap by providing a unified tool for simultaneously identifying clusters and directional connections. As demonstrated in the results, the new MARSS, as an integration of the MARSS and the clustering feature, is able to more accurately and robustly identify high-dimensional directional brain networks, compared to those approaches that identify connections only without the clustering feature. More importantly, despite patients' heterogeneous electrode arrays and underlying anatomic seizure foci, the new MARSS identified consistent network properties, including increases in the number of directional connections and the expansion of SOZ clusters after seizure onset. Future work will attempt to confirm our method's utility across larger samples.

We posit that the identification of clusters as well as directional connections of regions better reveals the interactions between regions that underlay seizure propagation. For example, the new MARSS accurately reflected the clinical history of Patient 1 (electrode placement shown in Figure \ref{fig:P1}) who tended to experience secondarily generalized seizures arising from a SOZ in the frontal lobe, as his SOZ cluster encompassed his entire EEG array (Figure \ref{fig:25aP1}). In contrast, the other 5 patients, whose electrographic seizures tended to remain focal, had complex partial seizures without secondary generalization. Accordingly, these 5 patients' SOZ clusters remained confined to focal electrodes (Figures \ref{fig:0aP2} and \ref{fig:25aP2}). Such results suggest an association between clinical symptoms of patients and the size of their SOZ clusters during seizure propagation. This association promises to be a key pursuit in future research, which will improve understanding of the
relationship between patients' epileptic brain networks and their medical impairments.

The SOZ and adjacent regions had highest directional connectivity changes at seizure onset. However, at other times (before seizure onset and several seconds after seizure onset), the directional connectivity of the SOZ did not differ from other regions (Figure \ref{fig:Sub3ADC}). Though existing network studies of interictal intracranial EEG data suggested that the SOZ had different connectivity properties as compared to other regions \citep{wilke2008estimation} during interictal phases, these studies were generally based on special interictal periods containing spikes or high-frequency oscillations \citep{korzeniewska2014ictal}. In contrast, our analysis was not based on these short interictal events that required selection using the appearance of epileptiform phenomena. Our result suggested that the directional connectivity of the SOZ on average was no different from other regions most of the time. Future work will evaluate directional connectivity during those short high-frequency interictal periods.

Although we believe the new MARSS provides unique insights on the pathophysiology of epileptic networks, the clinical importance of the new method lies in its ability--albeit preliminarily---to localize SOZs independently from the traditional interpretation of EEG for patients with focal epilepsy. Many network methods were developed to localize the SOZ, usually by comparing regions' connectivity strengths \citep{korzeniewska2014ictal,van2013ictal}. In contrast, the proposed SOZ localization method is a novel method of using the high-dimensional directional network results generated from the new MARSS, yielding great efficiency in detecting the unique directional connectivity property of the SOZ among many regions. More importantly, our SOZ localization method focuses on comparing regions' changes in directional connectivity at the seizure onset time. Thus, it is more sensitive to detect the network change due to seizure propagation  \citep{englot2016regional,Fisher05}.

Although the directional connectivity change of the SOZ at seizure onset was consistently among highest across all recorded regions for all 6 patients, selecting the region with the highest directional connectivity change at seizure onset alone does not always lead to accurate localization of the SOZ. This is because such SOZ localization depends on the accuracy of estimating the timing of seizure onset. Another potential problem is that a patient can have more than one region in the SOZ area. Our method addressed these limitations by examining directional connectivity changes both spatially and temporally for each region. In addition to selecting regions whose directional connectivity changes at the seizure onset time was among the top 10\% (spatially) across all the regions, we also excluded the regions whose maximum directional connectivity change (temporally) in interictal phases was 10\% larger than the region's directional connectivity change at the seizure onset time. The second step reduced false selections for SOZ localization. These two criteria for directional connectivity changes in the proposed SOZ-localization method ensured a high TPR and a low FPR for localizing the SOZ. Overall, our initial results show high concordance between the proposed method with the clinical ``gold standard" for patients with different epilepsy types and SOZ locations.

Despite flexibility and robustness of the new MARSS in identifying directional connections, the effectiveness of the proposed SOZ-localization method relies on clinical limitations of electrode coverage of the SOZ. In addition, since the new MARSS contains many parameters for a high-dimensional directional network, its high accuracies in SOZ localization also relies on sufficient data information, including recordings of at least three seizures per patient. Subsequent work will focus on generalizing our findings with a larger sample of patients with a range of surgical outcomes to determine accuracy and prognostic values.

While epilepsy surgery for the right patient can be transformative, epilepsy surgery remains underutilized because identification of the SOZ remains difficult and expensive, especially for patients without a clear clinical target \citep{Jacobs2012,rosenow2001presurgical,Surges2013}. Our statistical network analysis has the potential to provide an independent tool (without using clinical information) to localize the SOZ and to improve the clinical yield in determining the accuracy of SOZ localization.
\section{Appendix}

\subsection{Hypothesis testing on network changes at seizure onset}
This section explains how we evaluated the statistical significance of the increase in the number of directional connections  after seizure onset for the SOZ. We performed a hypothesis test for comparing the number of directional connections of the SOZ in the immediate postictal window (0-25 seconds after seizure onset) versus those in the interictal windows. For each patient, we first calculated the numbers of directional connections of the SOZ in all the interictal windows (25-second length, 1-second overlap) before seizure onset and used these numbers to create the null distribution for the number of directional connections of the SOZ. Against this null distribution, we obtained the p-value of the number of directional connections of the SOZ in the immediate postictal window. Similarly, we obtained the p-value of comparing the size of the SOZ cluster in the immediate postictal window versus those in the interictal windows. Our analysis showed that for all 6 patients, both increases in the number of directional connections and the size of the SOZ right after seizure onset were statistically significant (p-values $\leq$3\%).

\section{Data Availability}
Intracranial EEG data analyzed in the paper were collected from patients with epilepsy who received epilepsy diagnosis and treatment at the University of Virginia (UVA) Hospital. The UVA hospital owns all the intracranial EEG data. The third party can request the data directly from the UVA Health System. For questions regarding the data, please contact the co-author Mark Quigg via email at MSQ6G@hscmail.mcc.virginia.edu. 

\section*{Acknowledgments}
\noindent T. Zhang's research was funded by the NSF SES-2048991 and was supported in part by Research Computing at the University of Virginia and the University of Pittsburgh Center for Research Computing through the resources provided.

\bibliography{MARSS.bbl}

\bibliographystyle{plainnat}
\end{document}